# Economic Design of Memory-Type Control Charts:

# The Fallacy of the Formula Proposed by Lorenzen and Vance (1986)

**Amir Ahmadi-Javid**[1] and **Mohsen Ebadi**

Department of Industrial Engineering, Amirkabir University of Technology, Tehran, Iran

**Abstract.** The memory-type control charts, such as EWMA and CUSUM, are powerful tools for detecting small quality changes in univariate and multivariate processes. Many papers on economic design of these control charts use the formula proposed by Lorenzen and Vance (1986) [*Lorenzen, T. J., & Vance, L. C. (1986). The economic design of control charts: A unified approach. Technometrics, 28(1), 3-10, DOI: 10.2307/1269598*]. This paper shows that this formula is not correct for memory-type control charts and its values can significantly deviate from the original values even if the ARL values used in this formula are accurately computed. Consequently, the use of this formula can result in charts that are not economically optimal. The formula is corrected for memory-type control charts, but unfortunately the modified formula is not a helpful tool from a computational perspective. We show that simulation-based optimization is a possible alternative method.

## 1. Introduction

Statistical process control (SPC) plays a vital role in improving a firm's quality and productivity. Control charts are broadly-used tools of SPC for monitoring the quality of a production or service process. Designing a control chart means making appropriate decisions about the control chart parameters. The aim of economic design of a control chart is to determine the values of chart parameters in order to optimize an economic performance metric. The most popular metric is the long-run expected average cost, which were first studied by Duncan (1956) for Shewhart-type charts.

---

[1] Corresponding author's email address: ahmadi_javid@aut.ac.ir

Since then, many papers have studied economic design for various types of quality control charts (see for example, Chen, and Cheng (2007), Nenes, and Tagaras (2007), Wu, and Makis (2008), Ho and Quinino (2012), Yeong et al. (2013), and Liu et al. (2013)). For a review of the literature the readers are referred to Montgomery (1980), Keats et al. (1997), and Celano (2011).

Lorenzen and Vance (1986) claimed that for any given control chart they had found a unified formula to compute the long-run expected average cost. They stated

> *"A general process model is considered, and the hourly cost function is derived. This cost function simplifies when the recorded statistics are independent".* (Abstract, page 3, line 2)
>
> *"This article presents a general method for determining the economic design of control charts. The method applies to all control charts, regardless of the statistic used. It is only necessary to calculate the average run length of the statistic assuming the process is in-control and also assuming the process is out-of-control in some specified fashion. This is particularly easy when the statistics plotted are independent".* (Page 4, line 27)

This paper shows that these statements are not correct for memory-type control charts, such as EWMA-type, CUSUM-type, and Bayesian charts. Unfortunately, during the last three decades, several papers used Lorenzen and Vance's formula to study different problems on economic design of memory-type control charts; see, for example, Ho and Case (1992), Montgomery et al. (1995), Torng et al. (1995), Simpson and Keats (1995), Linderman and Love (2000a) and (200b), Love and Linderman (2003), Testik and Borror (2004), Carlos García-Díaz and Aparisi (2005), Chou et al. (2006), Yang and Sheu (2007), Serel and Moskowitz (2008), Lee (2010), Noorossana et al. (2014), Saghaei et al. (2014), Chiu (2015), Saniga et al. (2015), Ershadi et al. (2016), and Lu and Huang (2017). Most of these papers focused on economic design of EWMA-type control charts, and a few of them considered CUSUM charts.

The remainder of the paper is organized as follows. Section 2 briefly introduces the Lorenzen-Vance formula and develops a simulation method to accurately compute the long-run expected average cost for any given chart. Section 3 provides our numerical study for EWMA-type charts to show the incorrectness of using the formula developed by Lorenzen and Vance (1986) for memory-type control charts. Then, Section4 modifies the Lorenzen-Vance formula for memory-type charts. Finally, Section 5 concludes the paper.

## 2. Lorenzen-Vance formula and simlaution method

This section briefly introduces the general formula proposed by Lorenzen and Vance (1986), and describes our suggested simulation method in the following subsections.

### 2.1. Problem statement and Lorenzen-Vance formula

Consider a production process that continuously produces a product at constant rate and has two quality states: in-control and out-of-control. The process starts at the in-control state, under which its $q$ measurable quality characteristics follow a multivariate normal distribution $N_q(\boldsymbol{\mu}_0, \boldsymbol{\Sigma})$. The amount of time that the process stays in the in-control state before making a transition to the out-of-control state is stochastic and follows the exponential distribution with the mean $1/\lambda$. After a special cause occurs, the process mean shifts from $\boldsymbol{\mu}_0$ to $\boldsymbol{\mu}_1$. To control the process using a static control chart, a sample of size $n$ is taken at fixed sampling intervals every $h$ time units. Then, at the time epoch $mh$, $m = 1,2, ...$, a statistic $Y_\psi^m$ is computed and compared with a preset control limit $UL \in \Delta$, where, if applicable, $\psi \in \Psi$ includes all the specific designable parameter(s) of the control chart other than $UL$, $h$, and $n$. For some control charts, such as $\bar{X}$ and $T^2$ charts, there may be no specific designable parameter, while some others, such as MEWMA charts with equal and unequal exponential weights, may have only one or multiple parameters.

When the statistic of a chart exceeds the control limit, i.e., $Y^{mh} > UL$, the process is predicted to be out-of-control and a search for an assignable cause is initiated. Next, if any assignable cause exists, the signal is called a true alarm and a corrective action must be carried out in order to take the process back to the in-control quality state. Otherwise, the signal is called a false alarm and no action is done.

A *quality control cycle* (or cycle, for short) begins with the in-control state and continues until the occurrence, detection, and complete elimination of the assignable cause. Whenever an adjustment to the process is successfully made and the process is returned to the in-control state, a new cycle begins. At the beginning of each new cycle, the control chart is initialized as in the first cycle. Hence, the sequence of the cycles can be considered as a renewal stochastic process.

In the economic design of a control chart, the aim is to design some of the parameters $\psi$, $UL$, $h$, and $n$ such that an economic performance metric is optimized. Inspired by control theory, a widely-used objective is to minimize the long-run expected average cost defined by

$$F := \lim_{T \to \infty} E\left(\frac{\int_0^T C(t)dt}{T}\right) \tag{1}$$

where $C(t)$ denotes the instantaneous quality-control cost corresponding to the control chart specified by statistic $Y_\psi^m$, $m = 1, 2, ...$, and control limit $UL$. To have a well-defined problem, the above limit must finitely exist. The quality-control cost includes the production costs during the in-control and out-of-control periods, the cost of a false alarm, the cost of discovering assignable cause after a true alarm, the cost of repairing process after detecting an assignable cause, and the variable and fixed costs of sampling.

**Table 1.** Notation used in formula (3)

| | | | |
|---|---|---|---|
| $\boldsymbol{\mu}_0$ | Mean vector of $q$ process characteristics when process is in-control | $T_S$ | Expected time to sample and chart one item |
| $\boldsymbol{\mu}_1$ | Mean vector of $q$ process characteristics when process is out-of-control | $C_0$ | Production cost per time unit when the process is in-control |
| $\boldsymbol{\Sigma}$ | Covariance matrix of process | $C_1$ | Production cost per time unit when the process is out-of-control |
| $\lambda$ | Process failure rate | $C_{LR}$ | Cost for locating and repairing the assignable cause when one exists |
| $h$ | Length of each sampling interval | $C_F$ | Cost per false alarm which includes the costs of searching and testing for the cause |
| $n$ | Sample size | $a$ | Fixed cost of sampling |
| $T_L$ | Expected time to locate an assignable cause | $b$ | Variable cost of sampling |
| $T_R$ | Expected time to repair a detected assignable cause | $\gamma_1$ | A parameter that is 1 if production continues during search process, and 0 otherwise |
| $T_F$ | Expected search time for a false alarm | $\gamma_2$ | A parameter that is 1 if production continues during repair process, and 0 otherwise |

Let the random variables $CC$ and $CT$ denote the cost and time of a typical quality control cycle, respectively. As the cycles create a renewal stochastic process, the objective function $F$ can be given by

$$F = \frac{E(CC)}{E(CT)}. \qquad (2)$$

When the control chart is not memory-type, that is, statistics $Y_\psi^m$, $m = 1,2,...$, are independent, using the same method used by Duncan (1956), Ladany (1973), and Lorenzen and Vance (1986), based on (2) it can be shown that

$$F = \left\{ \frac{C_0}{\lambda} + C_1 \left[ -\tau + nT_S + \frac{h}{(1-\beta)} + \gamma_1 T_L + \gamma_2 T_R \right] + C_F s\alpha + C_{LR} \right.$$
$$\left. + \left[ \frac{a+bn}{h} \right] \left[ \frac{1}{\lambda} - \tau + nT_S + \frac{h}{(1-\beta)} + \gamma_1 T_L + \gamma_2 T_R \right] \right\} \qquad (3)$$
$$\div \left\{ \frac{1}{\lambda} + (1-\gamma_1) s\alpha T_F - \tau + nT_S + h/(1-\beta) + T_L + T_R \right\}$$

where

$$s := \frac{e^{-\lambda h}}{(1 - e^{-\lambda h})}, \tau := \frac{1 - (1 + \lambda h)e^{-\lambda h}}{\lambda(1 - e^{-\lambda h})},$$

and

$$\alpha := \Pr\{Y_\psi^{mh} > \text{UL} | \boldsymbol{\mu} = \boldsymbol{\mu}_0 \}, \beta := 1 - \Pr\{Y_\psi^{mh} > \text{UL} | \boldsymbol{\mu} = \boldsymbol{\mu}_1 \}$$

are the type-I and type-II error probabilities, which are the same for all $m = 1,2,...$. The other notation used above is given in Table 1. Note that it can alternatively be assumed that the process stops or continues during searching for an assignable cause, and repairing a detected cause depending on how the values of the two Boolean parameters $\gamma_1$ and $\gamma_2$ are set.

For a memory-type control chart, Lorenzen and Vance (1986) claimed that the formula (3) can be extended as follows:

$$F = \left\{\frac{C_0}{\lambda} + C_1[-\tau + nT_S + h(ARL_1) + \gamma_1 T_L + \gamma_2 T_R] + C_F \frac{s}{ARL_0} + C_{LR}\right.$$

$$+ \left[\frac{a+bn}{h}\right]\left[\frac{1}{\lambda} - \tau + nT_S + h(ARL_1) + \gamma_1 T_L + \gamma_2 T_R\right]\right\} \quad (4)$$

$$\div \left\{\frac{1}{\lambda} + \frac{(1-\gamma_1)sT_F}{ARL_0} - \tau + nT_S + h(ARL_1) + T_L + T_R\right\},$$

where $ARL_0$ and $ARL_1$ stand for the in-control and out-of-control ARLs (Average Run Lengths), respectively. When the statistics $Y_\psi^{mh}$, $m = 1,2,...$, are independent, the formula (3) can be retrieved from (4) by observing that $ARL_0$ and $ARL_1$ are equal to $1/\alpha$ and $1/(1-\beta)$, respectively.

In Lorenzen and Vance (1986), it is suggested that the ability to compute $ARL_0$ and $ARL_1$ is enough to apply the formula (4) even if the control-chart statistics at sampling epochs are dependent. For this purpose, three approximate methods can be used

(i) Using integral equations: An integral and a double-integral equation can be used to approximate the in-control and out-of-control ARLs, respectively (Rigdon, 1995a, b; Crowder, 1987).

(ii) Using Markov chains: A multistate Markov chain approximation that is obtained by a discretization method can be used to approximate the ARLs (Saccucci & Lucas, 1990; Runger & Prabhu, 1996; Woodall, 1984).

(iii) Using simulation: A simulation model can be used to approximate the ARLs (Lowrey et al., 1992; Linderman & Love, 2000b).

Because of the simplicity and generality, simulation is the most efficient method that has been used by almost all recent papers. The accuracy of the estimated ARLs using simulation depends on the number of simulation runs. As suggested by Lowrey et al. (1992), and Linderman and Love (2000b), performing 6,000 simulation runs provides good approximations for ARLs.

In Section 3, it will be demonstrated that, even if the ARLs are approximated very accurately through carrying out a very larger number of simulation runs, the formula (4) is incorrect for memory-type control charts. In Section 4, this formula is corrected.

## 2.2. Simulation method

Our alternative method to accurately estimate the objective function $F$ in (1) is to use a simulation model implemented in MATLAB (the simulation model is available online at the link https://www.dropbox.com/s/vva7yd3d8y0qqy2/SimulationCodeCostMEWMA.m?dl=0 for EWMA-type control charts[2]). Using this simulation model, for $N$ simulated cycles, the objective function $F$ in (1) can be approximated by

$$\hat{F} = \frac{\widehat{E(CC)}}{\widehat{E(CT)}} = \frac{\sum_{i=1}^{N} cc_i}{\sum_{i=1}^{N} ct_i} \tag{5}$$

where $cc_i$ and $ct_i$ represent the observed cost and time of the $i$th simulated cycle, respectively. Based on the strong law of large numbers, the estimated value $\hat{F}$ almost surely converges to $F$ whenever the variances of $CC$ and $CT$ are finite. Hence, the estimated value can be accurate up to any required level for sufficiently large $N$. The accuracy level of the simulation model versus $N$ will be discussed after presenting our numerical study in the next section.

## 3. The fallacy of the Lorenzen-Vance formula

To show that the Lorenzen-Vance formula (4) is not correct for memory-type control charts, it suffices to demonstrate it for a class of memory-type control charts. Here we consider EWMA and MEWMA control charts for which formula (4) has been extensively used by several papers under different settings. A similar analysis can be presented for other well-known memory-type control charts such as CUSUM and Bayesian charts, which is not given here for the sake of brevity. In the following, the EWMA-type control charts are briefly described, and then our numerical results will be represented based on the formula (4) and our simulation method explained in Section 2.2.

### 3.1 EWMA-type control charts

The EWMA control chart, developed by Lucas and Saccucci (1990), is one of the most commonly used memory-type control chart, which accumulates data from the past samples to detect small process shifts. Lowry et al. (1992) extended this chart to the multivariate EWMA (MEWMA) control chart for detecting mean shifts in multivariate processes. During the last two decades, considerable attention has been devoted to the statistical and economic design of EWMA-type control charts.

---
[2] This code can be used freely for personal use, or educational and research activities provided that the source is properly cited.

Consider the problem of monitoring $q$ quality characteristics over time. Let the random vector $\bar{X}_m, m = 1, 2, \ldots$, denote the mean statistic of the sample taken at the time epoch $mh$, which follows distribution $N_q(\boldsymbol{\mu}_0, n^{-1}\boldsymbol{\Sigma})$ when the process is in control. Let $0 < r_j \leq 1$, $j = 1, 2, \ldots, q$, be the exponential weight (or smoothing parameter) assigned to the past observations of characteristic $j$, and define the exponentially-weighted moving-average statistic

$$\mathbf{Z}_m = \mathbf{R}(\bar{X}_m - \boldsymbol{\mu}_0) + (\mathbf{I} - \mathbf{R})\mathbf{Z}_{m-1} = \sum_{j=1}^{m} \mathbf{R}(\mathbf{I} - \mathbf{R})^{m-j}(\bar{X}_m - \boldsymbol{\mu}_0), \quad m = 1, 2, \ldots \quad (6)$$

where $\mathbf{Z}_0 = \mathbf{0}$, $\mathbf{R} = \text{diag}(r_1, \ldots, r_q)$ is the diagonal matrix of exponential weights, and $\mathbf{I}$ is the identity matrix. The MEWMA chart signals a potential out-of-control process as

$$Y^m_{\psi=(r_1,\ldots,r_q)} = \left(\mathbf{Z}_m' \boldsymbol{\Sigma}_{\mathbf{Z}_m}^{-1} \mathbf{Z}_m\right)^{1/2} > UL, m = 1, 2, \ldots \quad (7)$$

where $UL > 0$ is the upper control-chart limit and where $\boldsymbol{\Sigma}_{\mathbf{Z}_m}$ is the covariance matrix of $\mathbf{Z}_m$.

If $r_j = r$, for all $j = 1, 2, \ldots, q$, then some calculations are simplified as

$$\mathbf{Z}_m = r\bar{X}_m + (1-r)\mathbf{Z}_{m-1} = \sum_{j=1}^{m} r(1-r)^{m-j}(\bar{X}_m - \boldsymbol{\mu}_0), \quad m = 1, 2, \ldots \quad (8)$$

and

$$\boldsymbol{\Sigma}_{\mathbf{Z}_m} = \left\{\frac{r[1-(1-r)^{2m}]}{n(2-r)}\right\} \boldsymbol{\Sigma} \quad (9)$$

The EWMA control chart is equivalent to the MEWMA by setting $q = 1$. For the special case of $r = 1$ ($q = r = 1$), the MEWMA (EWMA) chart is equivalent to Hotelling's $T^2$ chart (Shewhart's $\bar{X}$ chart). The quantity $ARL_1$ depends on the mean vectors $\boldsymbol{\mu}_0$, and $\boldsymbol{\mu}_1$, and covariance matrix $\boldsymbol{\Sigma}$. For $q = 1$, $ARL_1$ depends on $\boldsymbol{\mu}_1$ only through the non-centrality parameter $\delta$ defined by

$$\delta = \left(n(\boldsymbol{\mu}_1 - \boldsymbol{\mu}_0)' \boldsymbol{\Sigma}^{-1} (\boldsymbol{\mu}_1 - \boldsymbol{\mu}_0)\right)^{1/2}. \quad (10)$$

### 3.2. Evaluation of Lorenzen-Vance formula

As mentioned in Section 1, several papers explored the economic design of the EWMA and MEWMA charts using the Lorenzen-Vance formula (4). Therefore, we present our numerical study for the EWMA and MEWMA charts with equal exponential weights. This study compares the results obtained by the Lorenzen-Vance formula and our simulation method given in Section 2.2.

**Table 2.** Data of 36 benchmark instances

| Instance | $a$ | $b$ | $C_F$ | $C_{LR}$ | $C_0$ | $C_1$ | $T_S$ | $T_L + T_R$ | $\lambda$ | $\delta$ |
|---|---|---|---|---|---|---|---|---|---|---|
| $U_1\|M_1$ | 0.5 | 0.1 | 50 | 25 | 100 | 250 | 0.05 | 2 | 0.01 | 0.5 |
| $U_2\|M_2$ | 0.5 | 0.1 | 50 | 25 | 200 | 500 | 0.5 | 20 | 0.05 | 0.5 |
| $U_3\|M_3$ | 0.5 | 0.1 | 500 | 250 | 100 | 250 | 0.5 | 20 | 0.01 | 2 |
| $U_4\|M_4$ | 0.5 | 0.1 | 500 | 250 | 200 | 500 | 0.05 | 2 | 0.05 | 2 |
| $U_5\|M_5$ | 0.5 | 1 | 50 | 25 | 100 | 250 | 0.5 | 2 | 0.05 | 2 |
| $U_6\|M_6$ | 0.5 | 1 | 50 | 25 | 200 | 500 | 0.05 | 20 | 0.01 | 2 |
| $U_7\|M_7$ | 0.5 | 1 | 500 | 250 | 100 | 250 | 0.05 | 20 | 0.05 | 0.5 |
| $U_8\|M_8$ | 0.5 | 1 | 500 | 250 | 200 | 500 | 0.5 | 2 | 0.01 | 0.5 |
| $U_9\|M_9$ | 5 | 0.1 | 50 | 25 | 100 | 250 | 0.05 | 20 | 0.05 | 2 |
| $U_{10}\|M_{10}$ | 5 | 0.1 | 50 | 25 | 200 | 500 | 0.5 | 2 | 0.01 | 2 |
| $U_{11}\|M_{11}$ | 5 | 0.1 | 500 | 250 | 100 | 250 | 0.5 | 2 | 0.05 | 0.5 |
| $U_{12}\|M_{12}$ | 5 | 0.1 | 500 | 250 | 200 | 500 | 0.05 | 20 | 0.01 | 0.5 |
| $U_{13}\|M_{13}$ | 5 | 1 | 50 | 25 | 100 | 250 | 0.5 | 20 | 0.01 | 0.5 |
| $U_{14}\|M_{14}$ | 5 | 1 | 50 | 25 | 200 | 500 | 0.05 | 2 | 0.05 | 0.5 |
| $U_{15}\|M_{15}$ | 5 | 1 | 500 | 250 | 100 | 250 | 0.05 | 2 | 0.01 | 2 |
| $U_{16}\|M_{16}$ | 5 | 1 | 500 | 250 | 200 | 500 | 0.5 | 20 | 0.05 | 2 |
| $U_{17}\|M_{17}$ | 0.5 | 0.1 | 50 | 25 | 10 | 100 | 0.05 | 4 | 0.01 | 0.5 |
| $U_{18}\|M_{18}$ | 0.5 | 0.1 | 50 | 25 | 10 | 100 | 0.05 | 4 | 0.01 | 2 |

In Table 2, 18 adapted process scenarios with practical cost and process parameters are provided. The first 16 scenarios were based on Molnau et al. (2001), while the two others were taken from Montgomery et al. (1995). We created 36 instances based on these process scenarios which are denoted by $U_1|M_1$ to $U_{18}|M_{18}$. In all instances, $T_F = \gamma_1 = \gamma_2 = 0$. In both univariate and multivariate cases, $n = 1$, $h = 1.5$, and $UL = \sqrt{10.5}$. We also considered a univariate process with $\mu_0 = 0$ and $\sigma = 1$, and a trivarite process with $\boldsymbol{\mu}_0 = (0,0,0)'$ and covariance matrix

$$\Sigma = \begin{bmatrix} 2 & 1 & 1 \\ 1 & 3 & 1 \\ 1 & 1 & 3 \end{bmatrix}.$$

All runs were performed on a PC with Intel Core(TM)2 Quad CPU (Q8400), 2.66 GHz and 4 GB RAM. The average time to perform 100,000 simulation runs for each value of $r$ is about 974 seconds. The inverse relationship between run times and the magnitudes of the parameters $\delta$ and $\lambda$ has been observed in our numerical study. One can considerably improves run times by applying programming languages, such as C and FORTRAN, which are more efficient, but perhaps less user-friendly than MATLAB.

Tables 3 and 4 represent the results for univariate and multivariate cases, respectively. In these tables, for each scenario, a comparison is made between the average costs obtained from the simulation method and those computed from (4) for different values of the exponential weight $r$. For each value of $r$, rows S10 and S100 show the simulation-based optimization results based on $N = 10,000$ and $N = 100,000$ simulations run, respectively. In both Tables 4 and 5, to apply the Lorenzen-Vance formula, the in-control and out-of-control ARLs are determined based on 100,000 runs, which is very large compared to 6,000 runs considered by the papers recently used this formula.

In Tables 3 and 4, each row labeled %Dif gives the absolute of the difference percentage between the cost obtained by (4) and the simulation method for $N = 100,000$. These percentages, which increase up to 20%, clearly show that formula (4) is not correct. It can generally be seen that the difference increases as $r$ decreases. For instances $U_4|M_4$, $U_5|M_5$, $U_9|M_9$, $U_{16}|M_{16}$ and, $U_{18}|M_{18}$ the differences are larger, such that for $r = 0.05$ all of them are greater than 10%.

From both Tables 3 and 4, it can be seen that the average and the maximum of absolute (relative) differences between estimated cost values based on $N = 10,000$ and $N = 100,000$ simulation runs are 0.35 and 2.1 (0.002 and 0.018). This shows that $N = 10,000$ is sufficiently large to provide accurate estimated results using the simulation method.

When $r = 1$, the EWMA and MEWMA chart will be equivalent to their corresponding Shewhart-type control charts, i.e. $\bar{X}$, and $T^2$ chart, respectively. In this case, the control charts are not memory-type and the Lorenzen-Vance formula (4) is valid for them. Because in (4), we use estimated ARLs, the resulting numerical results are estimations whose errors tend to zero as $N$ becomes very large. By comparison of the values obtained from (4) and our simulation method with $N = 10,000$, it can be seen that they are very close with the maximum deviation of 0.35 %.

Using the exact results obtained by the formula (3), the row "%Error1" reports the relative errors of the values obtained by (4), which shows that the Lorenzen-Vance formula works correctly for the case of $r = 1$. Moreover, the row "%Error1" presents the relative errors of the values obtained by our simulation method, where the maximum error is less than 0.33 %. This observation can also be used to double-check the validity of the simulation method.

### 3.3. Inferior economic design with Lorenzen-Vance formula

This subsection evaluates how much using the Lorenzen-Vance formula can affect the final economic design. To this end, when the control limit $UL$ was fixed at $\sqrt{10.5}$, the economically-optimal values for $r$ were obtained by using both Lorenzen-Vance formula and simulation method to estimate the objective function in (1). The direct search is carried out over the values $r = 0.01, 0.02, ... ,1$ to find the optimal solution. Figure 1 depicts the optimal value of $r$ in the EWMA and MEWMA control chart for each one of the 36 instances given in Table 3. This figure discloses that the differences between the improper and proper optimal values based on Lorenzen-Vance formula and simulation method in all instances are significantly high, especially in instances $U_3|M_3$, $U_4|M_4$, $U_5|M_5$, $U_6|M_6$, $U_9|M_9$, $U_{10}|M_{10}$, $U_{15}|M_{15}$, $U_{16}|M_{16}$, and $U_{18}|M_{18}$.

Actually, considerable additional costs can be incurred when using the improper optimal values determined by the Lorenzen-Vance formula instead of the proper optimal values determined by the simulation-based optimization method. As demonstrated in Figure 2, the additional cost percentages vary from 0.20% to 45.64% for the EWMA chart, and range from 0.26% to 44.55% for the MEWMA chart.

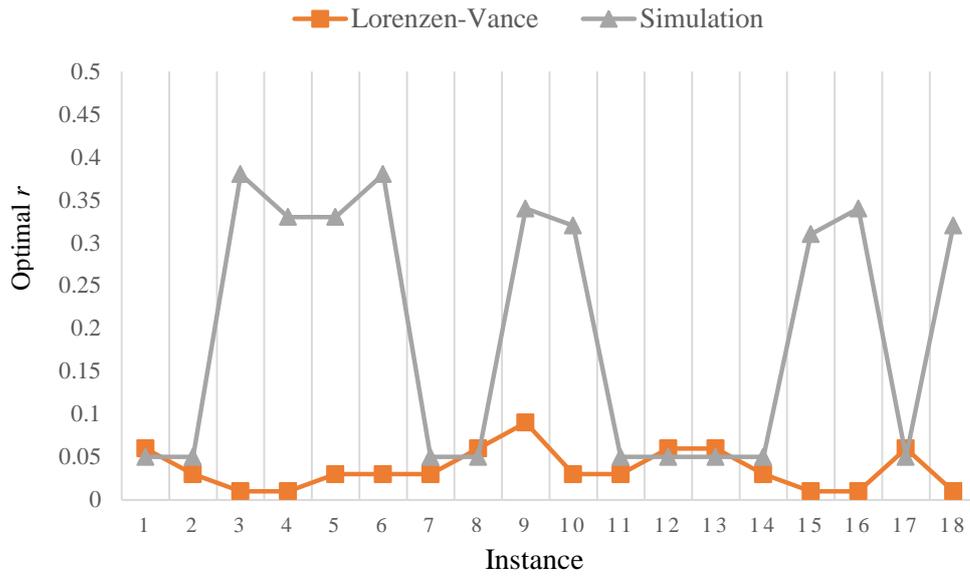

(a) EWMA, $q = 1$

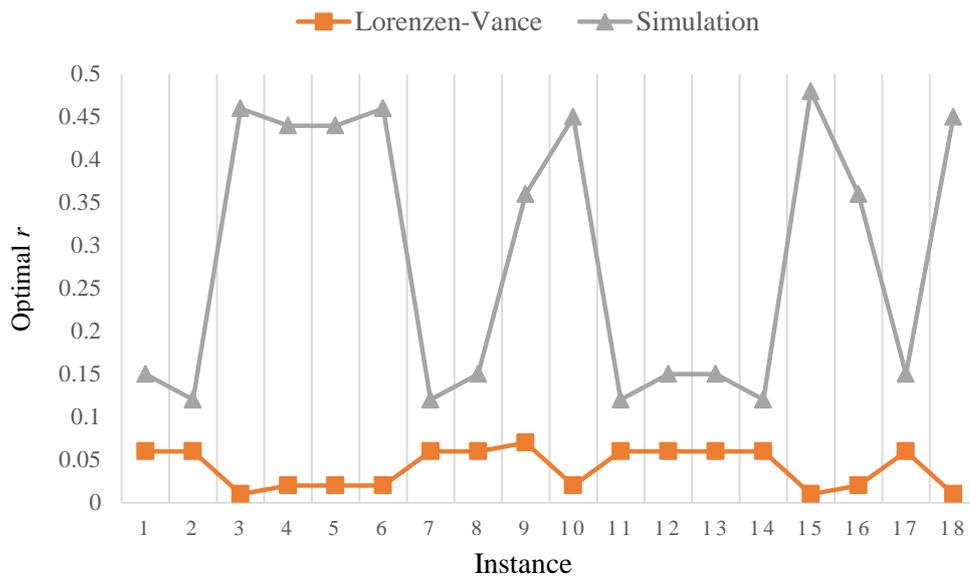

(b) MEWMA, $q = 3$

**Figure 1.** Comparison between improper and proper optimal values of $r$ based on Lorenzen-Vance formula and simulation method

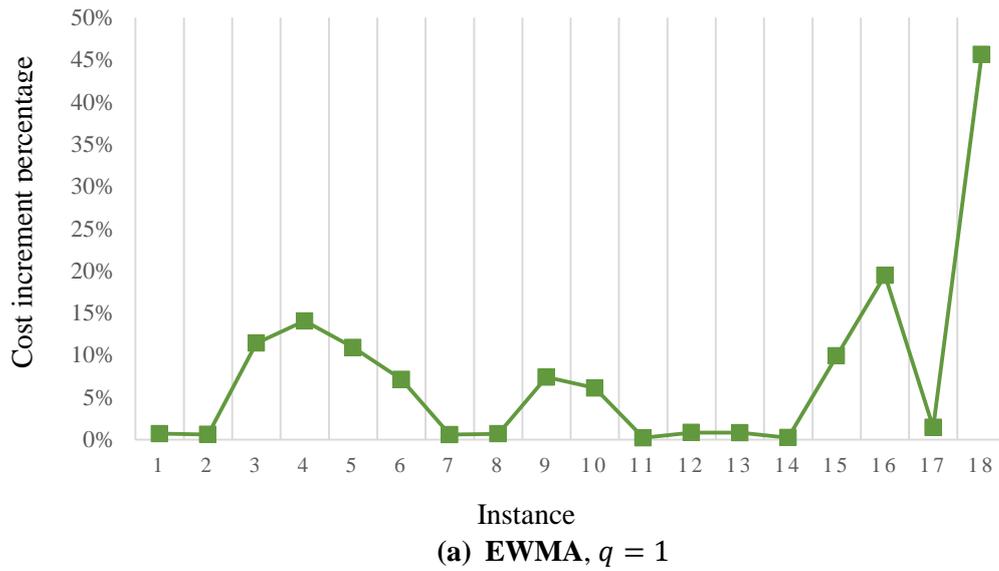

**(a) EWMA**, $q = 1$

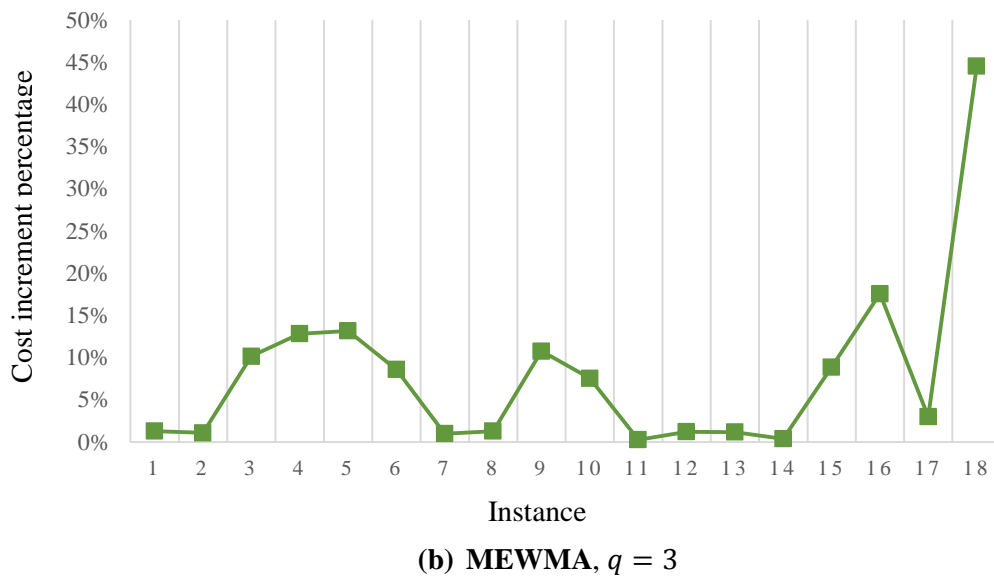

**(b) MEWMA**, $q = 3$

**Figure 2.** Cost increment percentage when Lorenzen-Vance formula is used instead of simulation method to determine optimal values of $r$

## 4. Modification of Lorenzen-Vance formula for memory-type control charts

This section shows how the Lorenzen-Vance formula (4) can be modified. The key point is that for a memory-type chart in-control and out-of-control ARLs at different time instances are not the same and depend on the samples taken before. Hence, we can modify the formula (4) as follows:

$$F = \left\{ \frac{C_0}{\lambda} + C_1[-\tau + nT_S + h(AARL_1) + \gamma_1 T_L + \gamma_2 T_R] + C_F ANFA + C_{LR} \right.$$

$$+ \left[\frac{a+bn}{h}\right]\left[\frac{1}{\lambda} - \tau + nT_S + h(AARL_1) + \gamma_1 T_L + \gamma_2 T_R\right]\bigg\} \quad (11)$$

$$\div \left\{\frac{1}{\lambda} + \frac{(1-\gamma_1)sT_F}{ARL_0} - \tau + nT_S + h(AARL_1) + T_L + T_R\right\}.$$

To obtain (11), $s/ARL_0$ and $ARL_1$ in (4) are replaced by $ANFA$ and $AARL_1$, respectively. The $ANFA$ represents the average number of false alarms per cycle. The $AARL_1$ is the average of $ARL_1^m$, $m = 1,2,...$, defined by

$$AARL_1 = \sum_{m=1}^{\infty} \Pr(A_m) \cdot ARL_1^m, \quad (11)$$

where $A_m$ is the event that the first assignable cause occurs between the $m$th and $(m+1)$st sample epochs, with

$$\Pr(A_m) = \exp(-m\lambda h) - \exp(-(m+1)\lambda h),$$

and where $ARL_1^m$, $m = 1,2,...$, denotes the out-of-control ARL given that $A_m$ happens.

To use (11) in practice, we may only consider the first $k$ terms of the summation in (12) if $k$ is chosen sufficiently large. One way to determine $k$ is to use the following criterion:

$$\sum_{m=1}^{k} \Pr(A_m) \geq 1 - \epsilon \Rightarrow k \geq \frac{-ln\epsilon}{\lambda h},$$

for some desired confidence level $0 < 1 - \epsilon < 1$. Note that as $m$ tends to $\infty$, $ARL_1^m$ tends to the steady-state out-of-control ARL. Hence, the ignorance of $ARL_1^m$, $m = k+1, k+2, ...$, is not problamaitic when $k$ is large enough. $ARL_1^m$, $m = 1,2,...,k$ and $ANFA$ can be approximated by simulation. The MATLAB simulation codes for calculating $AARL_1$ and ANFA for EWMA-type control charts are respectively available online at the following links[3]:

https://www.dropbox.com/s/xdqz4z4mw4m7qta/AARL1MEWMA.m?dl=0,

https://www.dropbox.com/s/kpr3nmert92xybv/ANFAMEWMA.m?dl=0.

---

[3] These codes can be used freely for personal use, or educational and research activities provided that the source is properly cited.

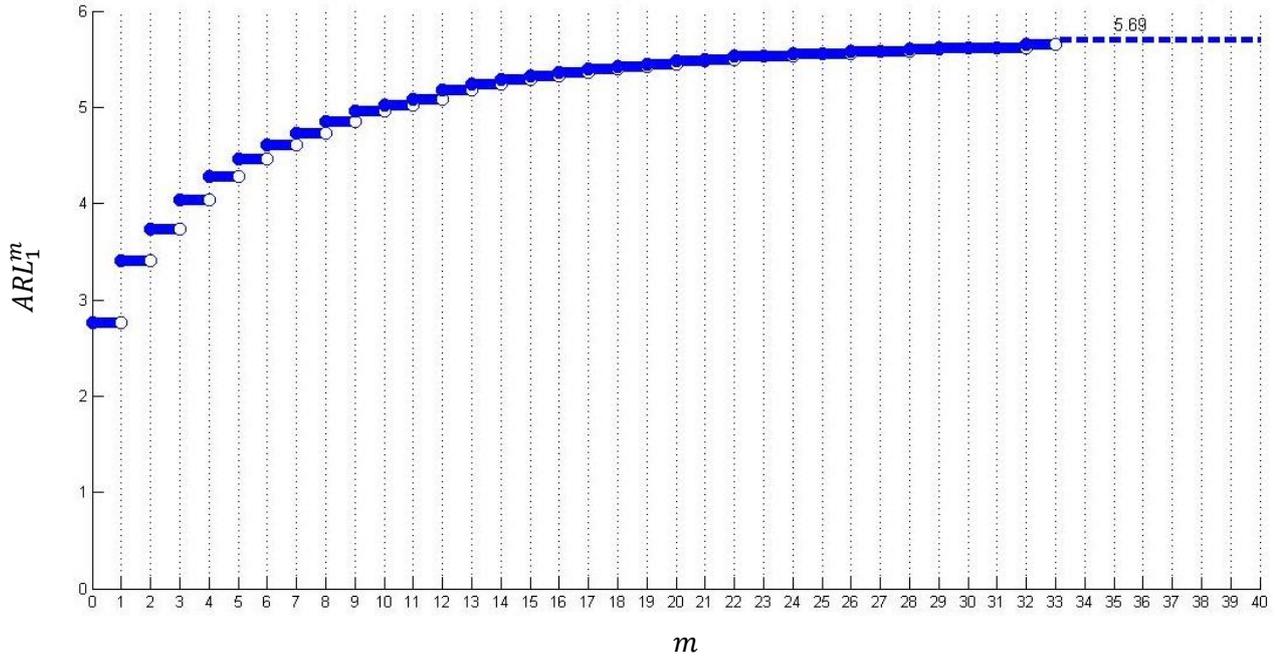

**Figure 3.** $ARL_1^m$, $m = 1, 2, \ldots, k$ for an EWMA control chart with $r = 0.05$, $\lambda = 0.05$, and $\delta = 2$

Figure 3 graphically displays $ARL_1^m$, $m = 1, 2, \ldots$, for the EWMA control chart with $r = 0.05$, $\lambda = 0.05$, and $\delta = 2$. These values converge to 5.69, which is the steady-state out-of-control ARL for this example. From this figure, it can clearly be seen that $ARL_1^m$, $m = 1, 2, \ldots$, are significantly different, and therefore, $AARL_1$ used in (11) is considerably greater than the out-of-control ARL used in (4), denoted by $ARL_1$ (which is identical to $ARL_1^1$).

To numerically check that the modified Lorenzen-Vance formula works correctly, we evaluate it on five instances $M_4, M_5, M_9, M_{16}$ and $M_{18}$ under the MEWMA chart with $r = 0.05$ or $r = 0.1$. Recall that the Lorenzen-Vance formula gives the most significant errors for these instances. Figure 4 depicts the values of $F$ obtained by the Lorenzen-Vance formula (4) (with $N = 10{,}000$), its modified version (11) (with $N = 10{,}000$ and $\epsilon = 10^{\wedge}(-10)$), and the simulation method (with $N = 100{,}000$). It can be seen that the results obtained from the modified formula (11) is considerably close to the true values obtained by the simulation method.

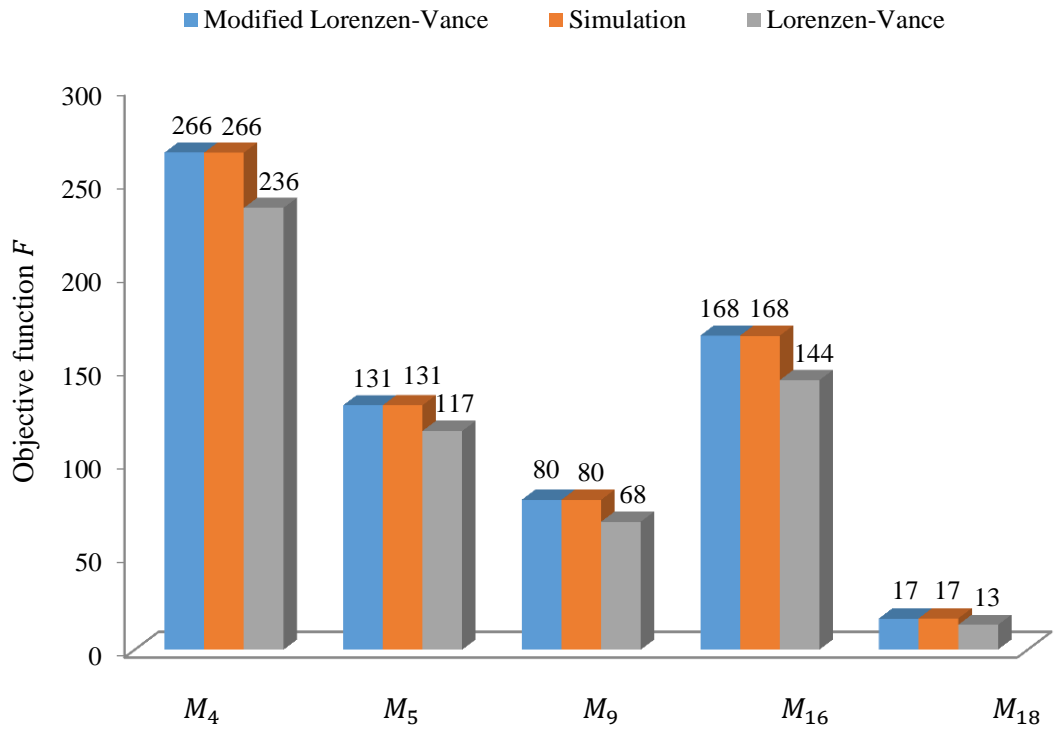

**(a)** $r = 0.05$

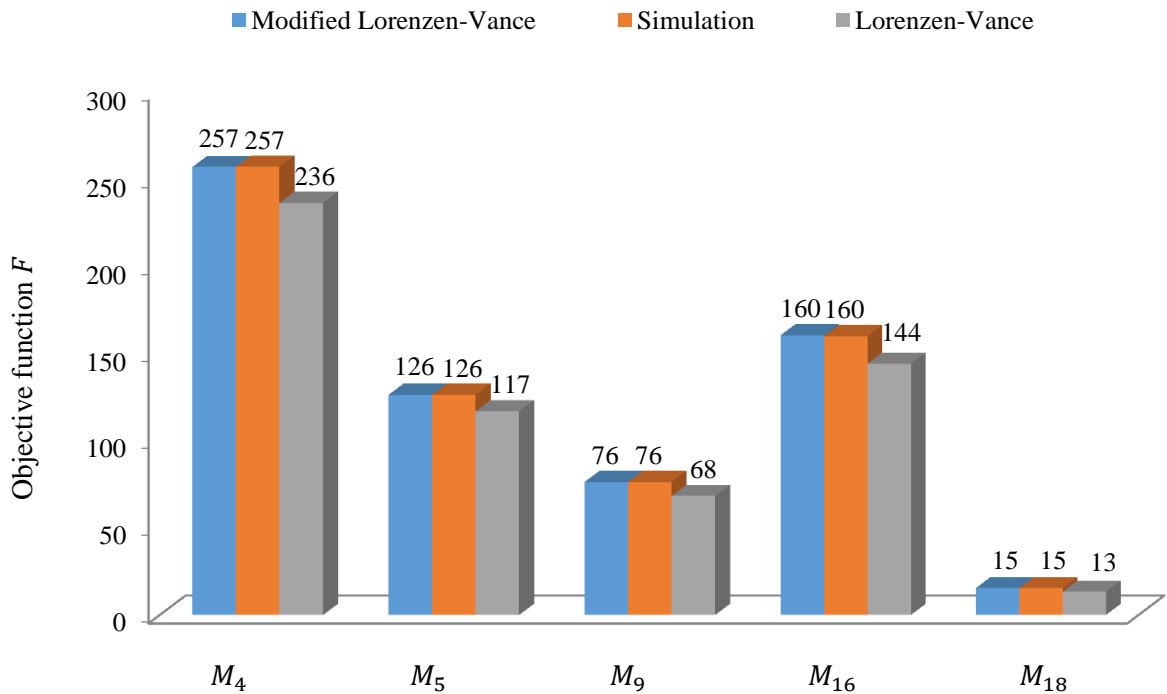

**(b)** $r = 0.1$

**Figure 4.** Comparison of the estimated values for the objective value $F$ obtained by modified Lorenzen-Vance formula, simulation method, and Lorenzen-Vance formula for instances $M_4, M_5, M_9, M_{16}$ and $M_{18}$ under the MEWMA control chart

Unfortunately, the modified formula (11) cannot be a basis for an efficient computational method. In fact, when using this formula, for a sufficiently large $k$ the quantities $ARL_1^m, m = 1, 2, \ldots, k$ must be computed by some method such as simulation so that $AARL_1$ can be approximated, which is very time consuming.

## 5. Conclusions

This paper shows that the classic formula proposed by Lorenzen and Vance (1986) cannot be used for memory-type control charts. It suggests using a simulation method where the accuracy level obtained by 10,000 simulation runs is satisfactory for both EWMA and MEWMA charts. Moreover, the paper emphasizes that the usage of the Lorenzen-Vance formula may result in very weak economic design of memory-type charts. Then, it modifies this formula by introducing new types of quality metrics.

The Lorenzen-Vance formula has commonly been used by many papers for 30 years in economic design of memory-type charts, especially EWMA-type charts. Hence, the results reported by these papers require reappraisal if they are based on the numerical studies conducted by applying this formula. This requires further investigation in future, which can use the simulation method proposed in this paper. Another interesting open research area is to propose a more efficient computational method rather than the simulation method proposed here.

**Table 3.** Comparison of results obtained by Lorenzen-Vance formula and simulation method for EWMA chart ($q = 1$)

| $r$ | Results | $U_1$ | $U_2$ | $U_3$ | $U_4$ | $U_5$ | $U_6$ | $U_7$ | $U_8$ | $U_9$ | $U_{10}$ | $U_{11}$ | $U_{12}$ | $U_{13}$ | $U_{14}$ | $U_{15}$ | $U_{16}$ | $U_{17}$ | $U_{18}$ |
|---|---|---|---|---|---|---|---|---|---|---|---|---|---|---|---|---|---|---|---|
| 0.05 | S10 | 156.65 | 342.72 | 97.47 | 270.09 | 134.10 | 189.79 | 174.05 | 315.40 | 83.46 | 223.97 | 214.19 | 286.03 | 144.71 | 419.39 | 116.14 | 172.77 | 44.29 | 17.02 |
| | S100 | 157.06 | 344.82 | 97.49 | 270.42 | 134.25 | 189.83 | 175.11 | 316.22 | 83.50 | 224.13 | 215.05 | 286.91 | 145.14 | 421.15 | 116.23 | 172.88 | 44.53 | 17.07 |
| | %Dif | 1.98% | 1.75% | 5.63% | 10.15% | 10.16% | 5.70% | 1.72% | 1.99% | 14.38% | 5.00% | 0.88% | 2.30% | 2.24% | 0.94% | 4.97% | 13.45% | 4.15% | 19.40% |
| 0.1 | S10 | 159.51 | 354.72 | 95.21 | 261.77 | 129.91 | 185.19 | 179.95 | 321.04 | 79.60 | 219.42 | 217.93 | 292.39 | 147.89 | 427.42 | 113.89 | 165.35 | 45.98 | 15.68 |
| | S100 | 160.55 | 354.91 | 95.32 | 261.56 | 129.80 | 185.36 | 180.05 | 323.11 | 79.35 | 219.52 | 217.98 | 294.32 | 148.86 | 427.52 | 113.96 | 164.88 | 46.60 | 15.71 |
| | %Dif | 0.77% | 0.57% | 3.43% | 7.11% | 7.10% | 3.43% | 0.57% | 0.79% | 9.92% | 3.00% | 0.27% | 0.93% | 0.90% | 0.28% | 3.03% | 9.26% | 1.58% | 12.42% |
| 0.20 | S10 | 171.98 | 380.75 | 93.83 | 253.55 | 125.80 | 182.41 | 192.68 | 345.70 | 75.69 | 216.22 | 225.15 | 318.56 | 161.02 | 443.09 | 112.29 | 157.84 | 53.37 | 14.74 |
| | S100 | 171.66 | 381.70 | 93.78 | 253.39 | 125.72 | 182.23 | 193.15 | 345.06 | 75.90 | 216.28 | 225.49 | 317.99 | 160.73 | 443.80 | 112.35 | 158.22 | 53.18 | 14.76 |
| | %Dif | 0.25% | 0.17% | 1.72% | 3.94% | 3.93% | 1.71% | 0.17% | 0.25% | 5.62% | 1.50% | 0.06% | 0.31% | 0.29% | 0.06% | 1.52% | 5.22% | 0.47% | 6.52% |
| 0.40 | S10 | 190.61 | 421.66 | 93.40 | 251.32 | 124.69 | 181.47 | 212.68 | 382.51 | 74.80 | 215.08 | 236.04 | 360.01 | 181.84 | 466.55 | 111.76 | 156.11 | 64.46 | 14.40 |
| | S100 | 191.10 | 420.71 | 93.29 | 250.91 | 124.49 | 181.22 | 212.20 | 383.49 | 74.80 | 215.20 | 235.58 | 360.88 | 182.28 | 465.61 | 111.81 | 156.10 | 64.75 | 14.44 |
| | %Dif | 0.12% | 0.01% | 0.61% | 1.47% | 1.47% | 0.61% | 0.01% | 0.12% | 2.10% | 0.51% | 0.00% | 0.11% | 0.11% | 0.00% | 0.51% | 1.95% | 0.21% | 2.25% |
| 0.60 | S10 | 205.70 | 444.48 | 94.08 | 257.18 | 127.63 | 182.80 | 223.80 | 412.38 | 77.86 | 217.10 | 241.30 | 394.13 | 198.99 | 477.94 | 112.78 | 161.96 | 73.48 | 15.00 |
| | S100 | 205.88 | 444.31 | 94.25 | 258.00 | 128.03 | 183.18 | 223.72 | 412.75 | 77.87 | 217.35 | 241.22 | 394.59 | 199.22 | 477.79 | 112.88 | 162.03 | 73.59 | 15.07 |
| | %Dif | 0.05% | 0.08% | 0.25% | 0.55% | 0.56% | 0.25% | 0.08% | 0.04% | 0.80% | 0.24% | 0.04% | 0.05% | 0.05% | 0.04% | 0.23% | 0.74% | 0.08% | 1.02% |
| 0.80 | S10 | 216.04 | 459.69 | 96.90 | 275.75 | 136.90 | 188.61 | 231.21 | 432.87 | 86.23 | 222.71 | 244.71 | 418.34 | 211.16 | 485.30 | 115.53 | 178.15 | 79.68 | 16.65 |
| | S100 | 216.28 | 459.37 | 96.95 | 276.40 | 137.22 | 188.71 | 231.05 | 433.33 | 86.11 | 223.03 | 244.66 | 418.93 | 211.46 | 485.20 | 115.70 | 177.94 | 79.82 | 16.74 |
| | %Dif | 0.07% | 0.01% | 0.04% | 0.20% | 0.20% | 0.05% | 0.01% | 0.07% | 0.19% | 0.02% | 0.00% | 0.06% | 0.06% | 0.00% | 0.02% | 0.18% | 0.11% | 0.08% |
| 1 | S10 | 223.32 | 469.29 | 102.78 | 308.83 | 153.47 | 200.66 | 235.88 | 447.29 | 102.03 | 234.69 | 246.86 | 435.97 | 220.04 | 489.92 | 121.45 | 208.71 | 84.05 | 20.17 |
| | S100 | 223.55 | 469.43 | 103.01 | 309.42 | 153.78 | 201.07 | 235.95 | 447.74 | 102.50 | 235.90 | 246.89 | 436.47 | 220.29 | 489.99 | 122.08 | 209.60 | 84.19 | 20.54 |
| | %Dif | 0.01% | 0.01% | 0.07% | 0.06% | 0.06% | 0.08% | 0.01% | 0.01% | 0.06% | 0.09% | 0.01% | 0.01% | 0.01% | 0.01% | 0.09% | 0.05% | 0.02% | 0.30% |
| | %Error1 | 0.02% | 0.11% | 0.00% | 0.00% | 0.00% | 0.00% | 0.01% | 0.02% | 0.00% | 0.00% | 0.00% | 0.02% | 0.02% | 0.01% | 0.00% | 0.00% | 0.03% | 0.00% |
| | %Error2 | 0.01% | 0.10% | 0.07% | 0.06% | 0.06% | 0.08% | 0.02% | 0.01% | 0.06% | 0.09% | 0.01% | 0.01% | 0.01% | 0.01% | 0.09% | 0.06% | 0.01% | 0.30% |

**Table 4.** Comparison of results obtained by Lorenzen-Vance formula and simulation method for MEWMA chart ($q = 3$)

| $r$ | Method | Instances | | | | | | | | | | | | | | | | | |
|---|---|---|---|---|---|---|---|---|---|---|---|---|---|---|---|---|---|---|---|
| | | $M_1$ | $M_2$ | $M_3$ | $M_4$ | $M_5$ | $M_6$ | $M_7$ | $M_8$ | $M_9$ | $M_{10}$ | $M_{11}$ | $M_{12}$ | $M_{13}$ | $M_{14}$ | $M_{15}$ | $M_{16}$ | $M_{17}$ | $M_{18}$ |
| 0.05 | S10 | 146.30 | 309.85 | 99.52 | 265.35 | 130.35 | 187.64 | 159.02 | 297.40 | 80.12 | 221.76 | 205.43 | 266.31 | 133.74 | 397.29 | 118.67 | 168.00 | 38.30 | 16.63 |
| | S100 | 145.72 | 310.59 | 99.70 | 265.68 | 130.64 | 187.47 | 159.31 | 296.30 | 80.08 | 221.92 | 205.65 | 265.48 | 133.29 | 397.95 | 119.06 | 167.77 | 37.95 | 16.70 |
| | %Dif | 3.44% | 3.67% | 7.80% | 11.11% | 10.52% | 5.52% | 4.02% | 4.01% | 14.60% | 4.94% | 2.38% | 4.44% | 3.76% | 2.12% | 7.22% | 14.15% | 8.05% | 20.67% |
| 0.1 | S10 | 143.31 | 305.26 | 97.74 | 257.30 | 126.34 | 183.44 | 156.74 | 291.54 | 76.43 | 217.57 | 204.20 | 260.31 | 130.71 | 394.58 | 116.93 | 160.85 | 36.54 | 15.42 |
| | S100 | 143.09 | 304.51 | 97.73 | 257.14 | 126.34 | 183.37 | 156.36 | 291.14 | 76.32 | 217.56 | 203.54 | 260.34 | 130.71 | 393.26 | 116.96 | 160.54 | 36.41 | 15.42 |
| | %Dif | 1.64% | 1.75% | 5.34% | 8.03% | 7.54% | 3.41% | 2.09% | 2.12% | 10.50% | 3.03% | 1.24% | 2.37% | 1.83% | 0.95% | 4.95% | 10.21% | 4.00% | 13.76% |
| 0.20 | S10 | 143.12 | 306.61 | 96.35 | 249.15 | 122.31 | 180.49 | 157.39 | 291.12 | 72.63 | 214.38 | 204.66 | 260.62 | 130.87 | 395.61 | 115.41 | 153.46 | 36.42 | 14.49 |
| | S100 | 143.10 | 307.16 | 96.35 | 248.83 | 122.17 | 180.50 | 157.65 | 291.14 | 72.75 | 214.48 | 204.52 | 260.43 | 130.75 | 395.42 | 115.46 | 153.67 | 36.41 | 14.52 |
| | %Dif | 0.58% | 1.07% | 3.07% | 4.56% | 4.24% | 1.80% | 1.28% | 0.90% | 5.94% | 1.57% | 0.80% | 1.04% | 0.69% | 0.62% | 2.83% | 5.80% | 1.48% | 7.61% |
| 0.40 | S10 | 146.70 | 319.34 | 95.60 | 244.08 | 119.71 | 179.03 | 163.57 | 298.20 | 70.61 | 212.76 | 208.21 | 268.10 | 134.61 | 403.71 | 114.59 | 149.63 | 38.53 | 14.01 |
| | S100 | 146.86 | 318.24 | 95.53 | 244.19 | 119.80 | 178.79 | 163.02 | 298.50 | 70.72 | 212.90 | 208.04 | 268.09 | 134.62 | 403.30 | 114.71 | 149.81 | 38.63 | 14.06 |
| | %Dif | 0.14% | 0.30% | 1.11% | 1.77% | 1.63% | 0.59% | 0.38% | 0.27% | 2.26% | 0.59% | 0.25% | 0.35% | 0.22% | 0.17% | 1.10% | 2.22% | 0.35% | 2.99% |
| 0.60 | S10 | 151.58 | 333.72 | 95.59 | 245.46 | 120.44 | 178.92 | 170.53 | 307.68 | 71.69 | 213.30 | 212.86 | 277.43 | 139.36 | 414.05 | 114.89 | 151.66 | 41.40 | 14.18 |
| | S100 | 151.76 | 331.73 | 95.82 | 246.37 | 120.90 | 179.39 | 169.58 | 308.06 | 71.66 | 213.31 | 212.10 | 277.76 | 139.52 | 412.37 | 114.91 | 151.61 | 41.51 | 14.18 |
| | %Dif | 0.18% | 0.14% | 0.45% | 0.73% | 0.69% | 0.27% | 0.17% | 0.21% | 0.98% | 0.21% | 0.06% | 0.19% | 0.16% | 0.03% | 0.39% | 0.94% | 0.41% | 1.02% |
| 0.80 | S10 | 156.80 | 344.77 | 96.88 | 253.89 | 124.71 | 181.65 | 175.88 | 317.92 | 75.34 | 215.29 | 215.87 | 287.87 | 144.62 | 420.86 | 115.83 | 158.64 | 44.49 | 14.75 |
| | S100 | 156.40 | 344.69 | 96.80 | 253.72 | 124.60 | 181.48 | 175.85 | 317.10 | 74.93 | 215.50 | 215.94 | 287.48 | 144.44 | 420.97 | 115.94 | 157.89 | 44.25 | 14.82 |
| | %Dif | 0.06% | 0.07% | 0.10% | 0.08% | 0.07% | 0.09% | 0.06% | 0.05% | 0.27% | 0.05% | 0.06% | 0.05% | 0.06% | 0.07% | 0.06% | 0.25% | 0.12% | 0.21% |
| 1 | S10 | 161.24 | 357.42 | 98.77 | 269.53 | 132.58 | 185.69 | 182.00 | 326.51 | 81.88 | 220.09 | 219.74 | 297.73 | 149.64 | 429.43 | 118.09 | 171.24 | 47.10 | 16.16 |
| | S100 | 160.65 | 356.02 | 99.00 | 269.21 | 132.42 | 186.05 | 181.34 | 325.39 | 81.83 | 220.30 | 219.22 | 296.37 | 148.94 | 428.28 | 118.25 | 171.13 | 46.76 | 16.22 |
| | %Dif | 0.15% | 0.06% | 0.02% | 0.08% | 0.08% | 0.03% | 0.07% | 0.15% | 0.02% | 0.01% | 0.05% | 0.14% | 0.14% | 0.05% | 0.00% | 0.02% | 0.31% | 0.01% |
| | %Error1 | 0.01% | 0.01% | 0.00% | 0.01% | 0.01% | 0.01% | 0.01% | 0.00% | 0.02% | 0.00% | 0.00% | 0.01% | 0.01% | 0.01% | 0.00% | 0.02% | 0.02% | 0.02% |
| | %Error2 | 0.16% | 0.07% | 0.02% | 0.07% | 0.07% | 0.04% | 0.07% | 0.16% | 0.04% | 0.01% | 0.05% | 0.15% | 0.15% | 0.06% | 0.00% | 0.03% | 0.33% | 0.03% |